\documentclass[prl,twocolumn,superscriptaddress,showpacs,amscd,amsmath,amssymb,verbatim]{revtex4}

\usepackage{graphicx}
\usepackage{textcomp}
\usepackage[OT1,T1]{fontenc}

\begin{document}
\newcommand{\kagome}{kagom\'e }

\title{Ground State of the Easy-Axis Rare-Earth Kagom\'e Langasite Pr$_3$Ga$_5$SiO$_{14}$}
\author{A. Zorko}
\affiliation{Laboratoire de Physique des Solides, Universit\'e Paris-Sud 11, UMR CNRS 8502, 91405
Orsay, France}
\affiliation{"Jo\v{z}ef Stefan" Institute, Jamova 39, 1000 Ljubljana, Slovenia}
\author{F. Bert}
\affiliation{Laboratoire de Physique des Solides, Universit\'e Paris-Sud 11, UMR CNRS 8502, 91405
Orsay, France}
\author{P. Mendels}
\affiliation{Laboratoire de Physique des Solides, Universit\'e Paris-Sud 11, UMR CNRS 8502, 91405
Orsay, France}
\author{K. Marty}
\affiliation{Institut N\'eel, CNRS and Universit\'e Joseph Fourier, BP 166, 38042 Grenoble, France}
\author{P. Bordet}
\affiliation{Institut N\'eel, CNRS and Universit\'e Joseph Fourier, BP 166, 38042 Grenoble, France}

\date{\today}
\begin{abstract}
We report muon spin relaxation ($\mu$SR) and $^{69,71}$Ga nuclear
quadrupolar resonance (NQR) local-probe investigations of the
kagom\'e compound Pr$_3$Ga$_5$SiO$_{14}$. Small quasi-static random
internal fields develop below 40~K and persist down to our base
temperature of 21~mK. They originate from hyperfine-enhanced
$^{141}$Pr nuclear magnetism which requires a non-magnetic Pr$^{3+}$ crystal-field
(CF) ground state. Besides, we observe a
broad maximum of the relaxation rate at $\simeq 10$~K which we
attribute to the population of the first excited magnetic CF level.
Our results yield a Van-Vleck paramagnet picture, at variance with
the formerly proposed spin-liquid ground state.
\end{abstract}
\maketitle

In magnetic systems, coupled spins are generally expected to
condense in an ordered state at low temperatures. Deviations from
this paradigm are found in systems possessing substantial
frustration, such as the celebrated geometrically frustrated kagom\'e
antiferromagnet. This corner-sharing triangular-based lattice indeed yields macroscopically degenerate spin configurations and tends to destabilize any N\'eel ordered state in favor
of a liquid phase. Experimental realizations have been until
very recently exclusively limited to transition-metal based
magnetism. For spins $S>1/2$, small perturbations to the purely
Heisenberg model, such as magnetic-anisotropy or minute
off-stoichiometry, were found to stiffen the spin system in an
ordered or glassy ground state~\cite{Bono,Wills}. Remarkably, in the $S=1/2$ case,
realized in the unique Herbertsmithite compound~\cite{Shores}, the
quantum fluctuations seem to help stabilizing the liquid phase~\cite{Helton,Mendels}
against such perturbation. The opposite limit
of the Ising kagom\'e lattice has been far less investigated due to the
scarcity of suitable systems. For large spins ($S>3/2$) the case of strong, yet finite, easy axis anisotropy has
been shown to be of particular interest. Beyond the Ising model on the kagom\'e
lattice, transverse quantum dynamics favor an unconventional
semi-classical spin liquid at low temperatures~\cite{Sen}. Besides, under applied magnetic field, a broad magnetization plateau is predicted~\cite{Sen2}.

The discovery of new members, RE$_3$Ga$_5$SiO$_{14}$
(RE = rare earth)~\cite{Bordet}, of the Langasite family has provided unique
realizations of the easy-axis kagom\'e antiferromagnet for RE=Nd, Pr.
Both Nd$_3$Ga$_5$SiO$_{14}$ (NGS) and Pr$_3$Ga$_5$SiO$_{14}$ (PGS) possess the same magnetic net, topologically equivalent to the kagom\'e lattice. The
magnetic anisotropy changes to easy-axis like at low temperature
(at 33~K in NGS and at 135~K in PGS). In NGS a
fluctuating ground state was evidenced down to
40~mK~\cite{ZhouNd,Zorko}, which remains to be fully
understood~\cite{Simonet}. PGS has been recently argued to be a spin liquid on
the verge of spin freezing, which could be induced by increasing the
chemical pressure \cite{ZhouPRL}. The spin-liquid ground state was proposed on the basis of the
absence of neutron magnetic Bragg peaks and a $T^2$ low-$T$ dependence of the
specific heat \cite{ZhouPr}. However, no neutron
diffuse scattering, characteristic of short-range correlations in
spin liquids, was observed. Further, when the magnetic field was
applied magnetic excitations \cite{ZhouPr} and spin dynamics
\cite{Lumata} were drastically affected.

In order to unambiguously determine the zero-field (ZF) ground state
of PGS, ZF local techniques -- muon spin
relaxation ($\mu$SR) and nuclear quadrupolar resonance (NQR) -- have been used. We
report the development of small random static magnetic fields below
$\sim 40$~K, which persist at least down to 21~mK. We propose that they
originate from hyperfine-enhanced $^{141}$Pr nuclear moments. This
implies that the crystal-field (CF) ground state of Pr$^{3+}$ ions is
\textit{non-magnetic} with a small energy gap to the first
magnetic CF level estimated to be 18(3)~K from relaxation
measurements. We stress that a non-magnetic ionic ground state is allowed
for the non-Kramers Pr$^{3+}$ ions ($J=4$), at variance
with the Kramers Nd$^{3+}$ ions ($J=9/2$). Our findings
contradict the formerly proposed spin-liquid picture \cite{ZhouPRL}. Instead, PGS should be regarded as a Van-Vleck
paramagnet.

\begin{figure}[b,t]
\includegraphics[trim = 1mm 15mm 1mm 15mm, clip, width=7.3cm]{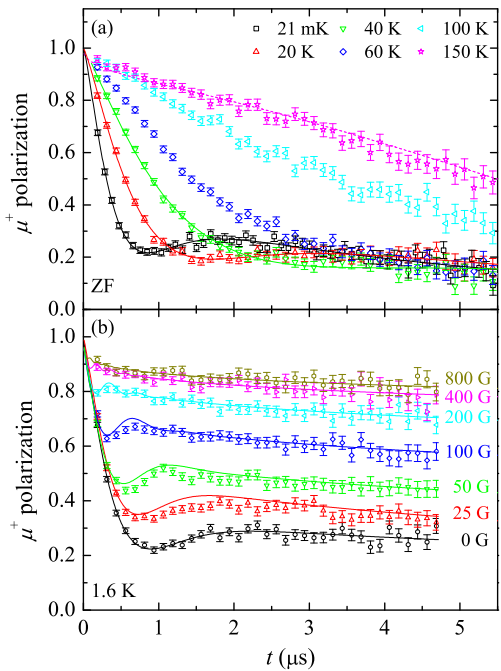}
\caption{(a) $T$-dependence of zero-field (ZF) $\mu^+$ depolarization. Solid lines are fits to the model $G_{VKT}\cdot \rm{exp}[-(\lambda t)^\alpha]$ with $\beta=1.3$, while dashed lines correspond to the Gaussian Kubo-Toyabe. (b) Longitudinal-field decoupling (see text) at 1.6~K. Solid lines show the agreement with the LF Voigtian Kubo-Toyabe model for the same $\beta=1.3$. The data has been corrected for muons stopping in the sample holder.}
\label{fig-1}
\end{figure}

$\mu$SR experiments were carried out on polycrystalline samples on the GPS and the LTF spectrometers at PSI,
Switzerland, and on the MuSR spectrometer at ISIS, England. The
samples were prepared by a solid state reaction. Their purity was
verified by x-ray powder diffraction and magnetization measurements.
$\mu$SR is well established for its unique sensitivity in detecting
local magnetic fields, determining their distributions and dynamics
\cite{Lee}. Muons implanted into a sample are initially almost 100\%
polarized along the beam direction and get depolarized in local
magnetic fields.

In Fig.~\ref{fig-1}(a) we show typical ZF $\mu^+$ relaxation curves,
which were measured in a broad $T$-range covering four
orders of magnitude. The relaxation gradually increases as the
temperature is lowered. Below $\simeq 2$~K it becomes
$T$-independent. In the ground state
the relaxation curve displays a marked dip at $t_{dip} \simeq 1$~$\mu$s [Fig.~\ref{fig-1}(b)]
followed by a slowly decaying tail at longer times which amounts to
about 1/3 of the full polarization. The 1/3-tail in powder-sample $\mu$SR is regarded as one of the firmest evidences of static magnetism while the lack of oscillations points to frozen disorder \cite{Lee}. In PGS muons thus experience a broad distribution of
randomly oriented static internal fields. Slow fluctuations that, nevertheless, persist down to the base
temperature (21~mK) give rise to a slow decay of the 1/3-tail.
The width of the static field distribution $\Delta/\gamma_\mu$ can
be estimated directly from the dip position, $\Delta/\gamma_\mu
\simeq 2/(\gamma_\mu t_{dip}) \simeq 20$~G, where $\gamma_\mu = 2\pi
\times 135.5$~MHz/T is the muon gyromagnetic ratio. Accordingly,
Fig.~\ref{fig-1}(b) shows that the muon relaxation is mostly
suppressed by applying, in the direction of the initial muon
polarization, an external field $\simeq 10$ times
larger than the internal random static fields. On the other hand,
the residual decay observed on the 1/3-tail in zero field demands
much higher fields to be suppressed, which confirms its dynamical
origin.

The low-$T$ magnetic state in PGS is therefore drastically
different from that in NGS, where ZF muon relaxation was monotonic
down to the lowest temperatures \cite{Zorko}. Moreover, we stress that the low-$T$ width $\Delta/\gamma_\mu \simeq
20$~G is neither typical for nuclear nor electronic field
distributions, the former usually being in the 1~G and the latter in the
1~kG range. For the cases of nuclear $^{141}$Pr magnetic moments
($\mu_I=4.25\mu_N$; $\mu_N$ is the nuclear magneton) and full
Pr$^{3+}$ electronic moments ($\mu_J = 3.57\mu_B$; $\mu_B$ is the
Bohr magneton) we calculated the dipolar-field-distribution width
\cite{foot2} at the three non-equivalent oxygen sites, in the
vicinity of which muons are most likely to reside. The corresponding values
for the nuclear and electronic fields are
$\Delta_I/\gamma_\mu=1.5-1.7$~G and
$\Delta_J/\gamma_\mu=2.3-2.7$~kG, respectively. We will address this
important issue that severely constrains the nature of the ground state in PGS later on.

\begin{figure}[t,b]
\includegraphics[trim = 1mm 5mm 1mm 10mm, clip, width=8.8cm]{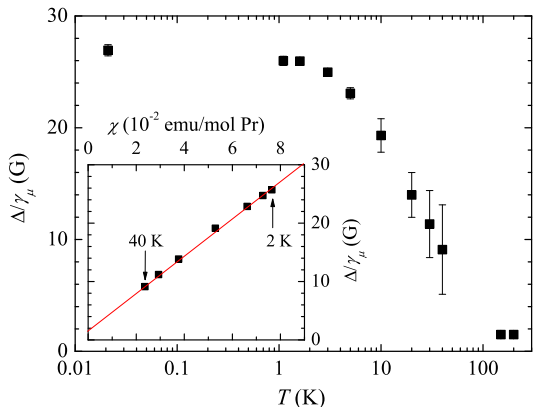}
\caption{Temperature evolution of the width of the static-random-internal-field distribution. The inset proves its linear scaling with bulk susceptibility measured in 10~G.}
\label{fig-2}
\end{figure}

In order to track accurately the $T$-dependence of the static
magnetism, the ZF data was fitted to the relaxation function
$G(t)=G_{VKT}(t)\cdot \rm{exp}[-(\lambda t)^\alpha]+G_0$
[Fig.~\ref{fig-1}(a)], where $G_0=0.07$ accounts for a constant fraction in our time window, the stretched
exponential term $\rm{exp}[-(\lambda t)^\alpha]$ accounts for dynamical $\mu^+$ relaxation causing the
decay of the 1/3-tail and $G_{VKT}(t)=\frac{1}{3}+\frac{2}{3} \left[
1-(\Delta t)^\beta) \right]{\rm exp} \left[ -(\Delta t)^\beta /\beta
\right]$ is the Voigtian Kubo-Toyabe relaxation expected for random static
fields with a distribution that interpolates between a Gaussian
($\beta=2$) and a Lorentzian ($\beta=1$) \cite{Crook}. This model fits nicely the
ZF data up to 40~K with a $T$-independent $\beta=1.3(1)$, which
indicates that the shape of the local field distribution does not
change with temperature. It is close to the Lorentzian distribution,
which is regularly the case in diluted canonical spin glasses
\cite{Uemura85}, but was observed also in magnetically dense spin
glasses \cite{Dunsiger,Wiebe}. The distribution shape may also be affected by the existence of at least 3
non-equivalent muon sites.
The VKT model was adapted to the LF case \cite{Lee} and fits equally well the LF data
taken at 1.6~K with fixed values of $\beta=1.3$ and
$\Delta/\gamma_\mu=26$~G [Fig.~\ref{fig-1}(b)]. The $T$-dependence of the distribution of
frozen fields, as fitted from the ZF data, is presented in Fig.~\ref{fig-2}. Above
150~K the relaxation is $T$-independent. As such, it
can be assigned to nuclear dipolar fields with a typical Gaussian
Kubo-Toyabe decay corresponding to $\Delta_I/\gamma_\mu=$1.5(3)~G, in agreement with our
calculation, $\Delta_I/\gamma_\mu=1.5-1.7$~G, for $^{141}$Pr nuclear
spins.

The static magnetism which develops below 40~K is
surprisingly weak, which likely justifies the failure of other less
sensitive techniques to detect it~\cite{ZhouPr,ZhouPRL,Lumata}. If
the frozen fields were to be ascribed to the Pr$^{3+}$ electronic
moments the concentration $c$ and/or the magnitude $\mu_e$ of the moments
would have to be strongly reduced. From the measured width $\Delta/\gamma_\mu$ of the
frozen field distribution and the relation $c\mu_{e} =\sqrt{2/\pi}
\mu_J \Delta/\Delta_J$~\cite{Uemura85}, we compute $c=0.007$ for
$\mu_e = \mu_J$. In principle, it is possible that the ground
state of the non-Kramers Pr$^{3+}$ ion in PGS is non-magnetic, except
at some rare sites where, because of the random Ga$^{3+}$/Si$^{4+}$ disorder on one of the Ga sites \cite{Bordet},
the CF would favor a magnetic ground state. This scenario, however, can be ruled out since
(i) PGS is an insulator so there exists no long-range
interaction that could induce a spin-glass state at 40~K for a
concentration of moments far below the site percolation threshold $c_p=0.65$
of the kagom\'e lattice. The main interaction is the
short-range exchange coupling, which is in praseodymium oxides
usually in the sub-Kelvin range~\cite{MacLaughlin}; (ii) In
the diluted electronic spin-glass scenario, one would
expect a Schottky peak at 170~mK~\cite{foot3} due to
the hyperfine splitting of the $^{141}$Pr nuclear levels, which is
not observed experimentally~\cite{ZhouPr}; (iii) The random Ga$^{3+}$/Si$^{4+}$ distribution
would not yield such a small concentration $c$.

We therefore propose an alternative scenario -- enhanced nuclear magnetism -- which is well
documented for materials based on non-Kramers rare earths with a non-magnetic
CF ground state and a strong hyperfine coupling
$A$~\cite{Bleaney, Teplov, AbragamBleaney}. Although these materials
are non-magnetic, they do possess a large Van-Vleck susceptibility
$\chi$ due to the proximity of low-lying
magnetic CF levels. The electronic shell can thus be polarized by
the nuclear magnetic moments through the hyperfine coupling and the
nuclear moments $\mu_I$ become effectively enhanced by a factor
$1+K=1+A\chi$ \cite{AbragamBleaney}. On the muon time scale, the
nuclear magnetism is static and disordered. This yields the usual
nuclear Kubo-Toyabe relaxation but with an enhanced width of the
field distribution $\Delta/\gamma_\mu=(1+A\chi)\Delta_I/\gamma_\mu$ as compared to the bare-nuclei width
$\Delta_I/\gamma_\mu$.

In PGS the field-distribution width indeed scales linearly
with bulk magnetic susceptibility below 40~K (inset to
Fig.~\ref{fig-2}). The width of
$1.8(2)$~G obtained by extrapolation to zero susceptibility is in agreement with the
1.5(3)~G value deduced from ZF data above 150~K
and with our calculations, 1.5-1.7~G. The slope
$A\Delta_I/\gamma_\mu=314$~G~mol/emu yields the hyperfine coupling
constant $A=174(20)$~emu/mol, in perfect
agreement with 187.7~emu/mol reported for Pr$^{3+}$
\cite{MacLaughlin,Bleaney}. The enhancement factor thus reaches the
value $K=15$ at low temperatures, very similar to other Pr-based
compounds \cite{MacLaughlin, Shu, MacLaughlinB}. Our ZF $\mu$SR
results can therefore be perfectly explained in the framework of the enhanced nuclear magnetism,
which proves that the ground CF state in PGS is non-magnetic. At higher temperatures, for $T\gtrsim \Delta_{CF}$, where $\Delta_{CF}$ is the gap to the first exited magnetic CF level, the Pr$^{3+}$ ions acquire a spontaneous fluctuating electronic moment. The hyperfine field is then motivationally narrowed in ZF and nuclear enhancement is suppressed \cite{Teplov}. Accordingly, our ZF data above 40~K shows a combination of static nuclear magnetism, observed above 150~K, and $T$-dependent electronic spin dynamics.

\begin{figure}[t,b]
\includegraphics[trim = 1mm 5mm 1mm 10mm, clip, width=8.8cm]{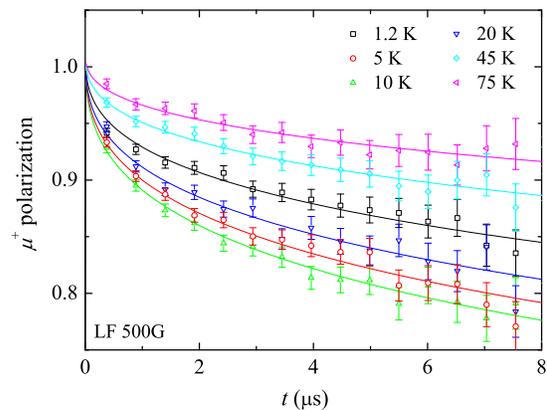}
\caption{$T$-dependent $\mu^+$ depolarization in a 500~G longitudinal field.}
\label{fig-3}
\end{figure}

\begin{figure}[t,b]
\includegraphics[trim = 1mm 2mm 1mm 8mm, clip, width=8.8cm]{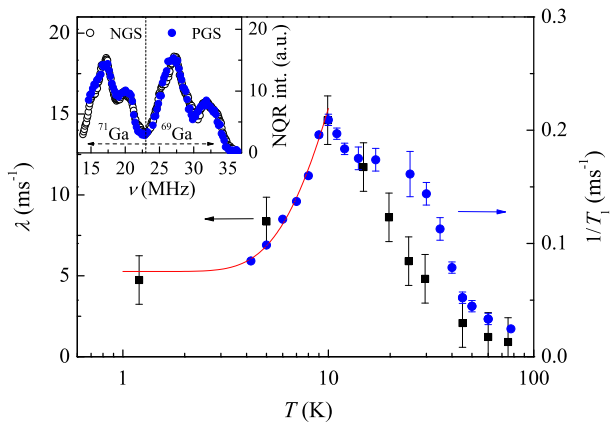}
\caption{$\mu^+$ relaxation rate $\lambda$ in LF 500~G (squares) and NQR spin-lattice relaxation rate $1/T_1$ measured at 27~MHz (circles). Solid line corresponds to the fit to the spin-gap model ($\Delta_{CF} = 18(3)$~K) with residual zero-temperatures relaxation. The inset shows comparison of the NQR spectra recorded in Pr$_3$Ga$_5$SiO$_{14}$ (PGS) and Nd$_3$Ga$_5$SiO$_{14}$ (NGS) at 80~K.}
\label{fig-4}
\end{figure}

Since static internal fields are small in PGS, one can easily
decouple muons from them by applying moderate longitudinal fields (500~G).
The remaining relaxation is then due to spin dynamics only. As shown in
Fig.~\ref{fig-3}, the relaxation was fitted satisfactory using a stretched exponential model with
a $T$-independent stretch exponent $\alpha=0.40(3)$. The
dynamical $\mu^+$ relaxation rate $\lambda$ exhibits a maximum
around 10~K (Fig.~\ref{fig-4}), which points to a substantial $T$-dependence of the magnetic fluctuations. The latter persist
down to the lowest temperatures, since the decay of the 1/3-tail in
ZF $\mu$SR is present even at 21~mK [Fig.~\ref{fig-1}(a)].

In the context where recent NMR experiments suggested that the magnetic fluctuations in
PGS are drastically affected by the applied field \cite{Lumata}, we performed a complementary ZF study using $^{69}$Ga ($I=3/2$) NQR. The $\mu$SR and NQR $T$-dependences of the relaxation rate are similar. In passing we note that
this similarity excludes the possibility that the $\mu^+$
charge has any appreciable effect on the near-neighbor Pr$^{3+}$ CF
levels in PGS, as observed in few other Pr-based compounds
\cite{Feyerherm,Tashma}.

The decrease of the NQR relaxation rate below
10~K can be fitted to the spin-gap expression $1/T_1=1/T_1^0 +
B\exp[-\Delta_{CF}/T]$ (Fig.~\ref{fig-4}), with $\Delta_{CF} = 18(3)$~K and a residual
relaxation $1/T_1^0=0.075$~ms$^{-1}$ for $T\rightarrow 0$. The gap
$\Delta_{CF}$ is in agreement with the inelastic neutron scattering
(INS) peak observed at 1.3-1.4~meV \cite{ZhouPr,Marty} and with CF
calculations \cite{Marty}. The INS peak is rather broad suggesting
distributed CF energy levels \cite{Marty}, which one can attribute
to random Ga$^{3+}$/Si$^{4+}$ disorder
\cite{Bordet}. This is reflected in a broad distribution
of local environments and leads to very broad $^{69,71}$Ga ($I=3/2$)
NQR spectra in PGS and NGS (inset to Fig.~\ref{fig-4}), instead of the
expected pair of narrow lines \cite{Zorko2}. This broad distribution
could be at the origin of the Voigtian $\mu^+$ decay instead of the expected
Gaussian one.

Both $\mu$SR and NQR measurements point to residual dynamics at
low $T$, extending down to at least 21~mK, as evidenced by the
dynamical decay of the ZF $\mu$SR polarization [Fig.~\ref{fig-1}(a)]. Since
the dynamics are present even for $T\simeq \Delta_{CF}/1000$, they are intrinsic to the non-magnetic ground state and not due to fluctuations between the
ground state and magnetic levels. The corresponding magnetic fluctuations rate
$\nu\sim 100$~kHz gives an estimate of the
$^{141}$Pr-$^{141}$Pr coupling $J_n=h\nu\sim5$~$\mu$K. This indirect
nuclear coupling is mediated by the electronic coupling
\cite{Bleaney} $J_e=J_n(g_J\mu_B I/K\mu_I)^2\sim 15$~mK. The latter
admixes excited CF states into the ground CF singlet,
which otherwise has a quenched total angular momentum in isolated ions.
This is not in contradiction with the Van-Vleck paramagnet picture,
because in PGS $J_e/\Delta_{CF}$ is far below the critical value
which would allow ground-state moments to form spontaneously
\cite{Goremychkin}.

In conclusion, we have found a very weak quasi-static magnetism in PGS which originates from
hyperfine-enhanced $^{141}$Pr nuclear magnetism. This enhancement unambiguously assigns PGS to be a
Van-Vleck paramagnet, which excludes the possibility of a collective
spin-singlet ground state. Much alike the rich family of pyrochlores, Langasites seem to present a variety of physical behaviors associated with the nature of the rare-earth ion. Our study calls for future indepth investigations of the single-ion properties of other members of the family, e.g., spin liquids should be found for Kramers ions with potentially enhanced exchange coupling. The Langasite family certainly opens the way to a new confrontation between theory and experiments on kagom\'e lattices with strong local anisotropy. Finally, our study serves as the zero-field basis to understand the surprising development of
short-range spin correlations in PGS at much higher temperatures under an applied field \cite{ZhouPr}.

We acknowledge technical assistance of A. Amato, C. Baines and A.
Hillier in $\mu$SR measurements and fruitful discussions with R. Ballou and V. Simonet. The work was supported by the EC
Marie Curie Grant No. MEIF-CT-2006-041243, by the ESF HFM Exchange
Grant No. 2292 and by the ARRS project No. J1-2118.

\appendix


\begin{thebibliography}{000}
\bibitem{foot3} The Pr$^{3+}$ hyperfine coupling of 1.093~GHz \cite{Bleaney} predicts internal fields of 335~T at the nucleus for full electronic moments and leads to the Schottky peak at 170~mK. Its height for the concentration $c=0.007$ of Pr$^{3+}$ moments is well above experimental values \cite{ZhouPr}.
\label{foot3}
\end{thebibliography}
\end{document}